\documentstyle[preprint,floats,prl,aps,epsf]{revtex}
\begin{document}

\preprint{\vbox{\hbox{Revised}\hbox {October 2003} }}

\draft
\title{MSSM from AdS/CFT} 
\author{\bf Paul H. Frampton$^{(a)}$ and Thomas W. Kephart$^{(b)}$}
\address{(a)Department of Physics and Astronomy,\\
University of North Carolina, Chapel Hill, NC  27599.}
\address{(b)Department of Physics and Astronomy,\\
 Vanderbilt University, Nashville, TN 37325.}
\maketitle
\date{\today}

\begin{abstract}
We develop a robust version of the MSSM from a $Z_{12}$ 
orbifolded AdS/CFT, with natural low scale unification 
and briefly discuss cosmology in this testable model.
\end{abstract}
\maketitle

\newpage

\bigskip
\bigskip

\noindent {\it Introduction}

\bigskip

Our collective understanding of particle physics is now called the 
standard model. Similarly, in
cosmology, the sum of our knowledge is contained in the standard 
cosmological model (or standard
big bang model). What these two standard models have in common is their 
lack of derivation from
fundamental principles. String and M theory provide hope we can 
understand both the micro world and the macro world
at a much deeper level and replace the standard models with standard 
theories. Here we present
a modest step toward such a synthesis.

\noindent While the standard model (SM) was still in its infancy, the theoretical
observations of the crossing of the running gauge couplings \cite{GQW} as
well as other facts led to attempts to unify the SM into one grand theory 
\cite{GG}. Although minimal SU(5) has been excluded experimentally, variants
and extensions thereof \cite{FG} remain viable. Low-energy supersymmetry is
natural in models derived from string theory, improves the accuracy of
unification of the couplings \cite{ADF,ADFFL} and raises the GUT scale to $%
\sim 10^{16}$ GeV.

Recent progress in string theory\cite{M,GKP,W} suggests a different path to unification
where the standard gauge group $SU(3)_{C}\times SU(2)_{L}\times U(1)_{Y}$ is
embedded in a semi-simple product gauge group $G=\Pi _{i}G_{i}$, reminiscent
of the Pati-Salam\cite{PS} and Trinification\cite{trinification} models, as inspired by gauge theories
arising from compactication of the IIB superstring on an orbifold. In such
quiver gauge theories the unification of couplings happens not by the
logarithmic evolution over a enormous desert covering many orders of
magnitude in energy scale, but occurs abruptly at a much lower scale $M_{U}$
through the diagonal embeddings of $SU(3)_{C} \times SU(2)_L \times U(1)_{Y}$ in 
$G$. If $M_{U}$ is sufficiently low (say $\sim $TeV) then the theory becomes
testable, since additional highly constrained and patterned particle content
is required at $M_{U}$.

Our focus is on particle physics models with measurable consequences,
derivable from orbifolded $AdS\otimes S^{5}, i.e.,$ $AdS_{5}\times %
S^{5}/\Gamma $ where $\Gamma $ is a finite group. Viable models of this type
are rare\cite{F1,F2} because of the numerous constraints imposed on the spectrum. These
include: (1) bifundamental matter due to orbifolding, (2) TeV unification to
allow experimental tests at accelerators, (3) consistent values for $\alpha
_{S},$ $\alpha $ and sin$^{2}\theta _{W},$ (4) three families, and (5) low
energy $\mathcal{N}$ = 1 supersymmetry. The two models we will present arise
from the only known abelian $Z_{n}$ orbifolded $AdS\otimes S^{5}$ with all
these desirable traits. We believe that it will be difficult to find many
more such models, and our work leads us to conclude that it is very unlikely
to find even one more with $n\leq 12.$ Hence, the scarcity of TeV scale
unification models derivable from string theory is strong motivation to
investigate the two models presented here. We would like to have as complete
a model as possible, and this means inclusion of gravity. However, gravity
in general, and cosmology in particular, within the $AdS/CFT$ scenario is a
delicate issue. In order to introduce gravity, conformal symmetry must be
broken and this must be provided by corrections to the basic field theoretic
model. This is an important topic, but we defer further discussion until we
have introduced our models.

\noindent We will show that the standard model may be unified at a scale of
approximately 4 TeV in a gauge theory based on the group $SU(3)^{12}$. There
are two such models, and in each the gauge hierarchy problem is ameliorated.
At the same time the models predict $\alpha _{3}(M_{Z})$ and $sin^{2}\theta
(M_{Z})$ both fall within experimental bounds. Finally, the models predict
the existence of precisely three chiral families of quarks and leptons, and
possesses ${\cal N}=1$ supersymmetry above the 4 TeV scale. At this scale
the models become distinguishable, but below 4 TeV we can still treat their
renormalization group behaviors the same.

We assume that supersymmetry is broken at the unification scale, so, in a
bottom-up approach, we should examine the running of all three of the SM
couplings with length scale $\mu $ as discussed in \textit{e.g.} 
\cite{PHF,Frampton:2003cp}. 
At the Z-pole\cite{PDG} $\alpha _{Y}(M_{Z})=0.0101,\alpha _{2}(M_{Z})=0.0338,\alpha
_{3}(M_{Z})=0.118\pm 0.003$ (the errors in $\alpha _{Y}(M_{Z})$ and $\alpha
_{2}(M_{Z})$ are less than 1\%) and run between $M_{Z}$ and $M$ according to
the SM equations 
\begin{eqnarray}
\alpha _{Y}^{-1}(M) &=&(0.01014)^{-1}-(41/12\pi )y  \nonumber \\
&=&98.619-1.0876y  \label{Yrun}
\end{eqnarray}
\begin{eqnarray}
\alpha _{2}^{-1}(M) &=&(0.0338)^{-1}+(19/12\pi )y  \nonumber \\
&=&29.586+0.504y  \label{2run}
\end{eqnarray}
\begin{eqnarray}
\alpha _{3}^{-1}(M) &=&(0.118)^{-1}+(7/2\pi )y  \nonumber \\
&=&8.474+1.114y  \label{3run}
\end{eqnarray}
where $y=\mathrm{ln}(M/M_{Z})$.

From Eqs.(\ref{Yrun},\ref{2run}) we find the scale\cite{PP,PF,DK} where $\mathrm{sin}%
^{2}\theta (M)=\alpha _{Y}(M)/(\alpha _{2}(M)+\alpha _{Y}(M))$ satisfies $%
\mathrm{sin}^{2}\theta (M)=1/4$ to be $M\simeq 4$ TeV. Furthermore, from
Eqs.(\ref{2run},\ref{3run}), the ratio $R(M)\equiv \alpha _{3}(M)/\alpha
_{2}(M)$ is $R(M)\simeq 3.5$, 3, 2.5, and 2 correspond to $M=M_{Z}$, $%
M_{3},M_{5/2},M_{2}\simeq 91\mathrm{GeV,}$ $400\mathrm{GeV},~~4\mathrm{TeV},%
\mathrm{and}~~140\mathrm{TeV}$ respectively. The proximity of $M_{5/2}$ and $%
M$, accurate to a few percent, suggests strong-electroweak unification at $%
M_{U}\simeq 4$ TeV. This scale will be taken as input in the model presented below.

Since the required ratios of couplings at $M_{U}\simeq 4$ TeV is: $\alpha
_{3}:\alpha _{2}:\alpha _{Y}::5:2:2$ it is natural to examine $Z_{12}$
orbifolds with CFT gauge groups $SU(3)^{12}$ (There are a large number of
such theories\cite{Kephart:2001qu,KephartPaes}, but the constraints (1) through (5) above
eleminates all but one orbifold choice.) and diagonal embeddings of Color
(C), Weak (W) and Hypercharge (H) in $SU(3)^{2},SU_{W}(3)^{5},SU_{H}(3)^{5}$
respectively. [To be precise, the hypercharge must come partially ($\frac{1}{%
3})$ from the same diagonal $SU(3)$ of $SU_{W}(3)^{5}$ where $SU(2)_{L}$
arises, and partially ($\frac{2}{3})$ from the diagonal $SU(3)$ of $%
SU_{H}(3)^{5}.$ As both diagonal $SU(3)$s arise from $SU(3)^{5}$s, the
ratios are as above.]

Both our ${\cal N}=1$ models arise from the same orbifold choice and will
have a top-down construction starting from the AdS/CFT correspondence \cite
{M,GKP,W} which suggests that the model satisfies conformality at an
infra-red fixed point; in fact, for the present model this appears even more
likely than in \cite{PHF} because ${\cal N}=1$ supersymmetry implies
the presence of non-renormalization theorems.

\bigskip \bigskip

\newpage

\noindent \textit{Description of the Model}

\bigskip

Maintaining ${\cal N}=1$ supersymmetry at the 4 TeV scale, requires an
embedding of the abelian finite orbifolding group $Z_{12}$ in the SU(4)
isotropy of the 5-sphere of $AdS_{5}\times S^{5}$ of type \textbf{4} = $%
\{a_{\mu }\}$ with $a_{\mu }=(i,j,k,0)$ where $a_{\mu }$ is a shorthand for $%
exp(2\pi ia_{\mu }/12),$ and $i, j,$ and $k$ are integers between 1 and 11.

Out of the 10 possible choices for an ${\cal N}=1$ embedding we must
choose a$_{\mu }=(1,2,9,0)$ if we are to agree with constraints (1), (2),
(3), and (5). This choice of $a_{\mu}$ is necessary to have three chiral families and to 
have simultaneously a scalar sector
which allows breakdown from $SU(3)^{12}$ to the standard gauge group.
Constraint (4) then restricts the embedding of $SU(3)_{C}%
\times SU(2)_{L}\times U(1)_{Y}$ in $SU(3)^{12}$ and the allowed patterns of
spontaneous symmetry breaking to arrive at $SU(3)_{C}\times SU(2)_{L}\times %
U(1)_{Y}$ from $SU(3)^{12},$ since changing the embedding changes the
arrangement of scalar fields relative to the SM gauge group. Note, the
corresponding $\mathbf{6=(4}\times \mathbf{4})_{A}=(1,2,3,-3,-2,-1)$ is real
as required by consistency of the theory \cite{FK}, and this provides yet
further constraint.

At first sight there is too much arbitrariness in identification of the
quiver nodes as C, W or H. The twelve nodes must be identified as two C's,
and five each of W and H. (The notation refers to the intermediate
trinification $SU(3)_{C}\times SU(3)_{W}\times SU(3)_{H}.$) The numbers of
C, W, H nodes is dictated by prediction of the correct $\alpha _{3}(M_{Z})$
and $sin^{2}\theta (M_{Z})$. However, there is only one ambiguity,
as we will show
there are only two possible embedding assignments. Consider first the choice
of two C nodes. There are six inequivalent choices where the two C's are
separated by between 1 and 6 places on the dodecagonal quiver. Now consider
the positioning of the W's and H's. For each assignment of the two C's there
are precisely 252 ways of assigning W's and H's: this is the number of
unordered partitions of the integer 6. This degeneracy is almost completely
removed by the requirements of spontaneous symmetry breaking to the $%
3_{C}3_{W}3_{H}$-model and then to the standard model, where we require
there are exactly three chiral families. In counting families it is
convenient to bear in mind anomaly cancellation (which is guaranteed by the
construction). This means that under $3_{C}3_{W}3_{H}$ the only
bi-fundamental combination that can occur is 
\[
(3,\bar{3},1)+(\bar{3},1,3)+(1,3,\bar{3})
\]
and so it is sufficient to count the $(3,\bar{3},1)$'s.

The first observation is that the two C's cannot be separated by more than 3
places on the dodecagon because the complex scalars available to break the $%
SU(3)\times SU(3)$ to the required $SU(3)_{C}$ diagonal subgroup do not
exist for these cases.

Suppose that the two C's are next-to-nearest neighbors. Then by considering
all the ten possible nodes among which to distribute the five W's one can
easily see that not more than two chiral families are possible.

Now assign the two C's to be neighboring nodes. The
VEVS of the scalars must be able to break the sets of five W's and five H's
to their respective diagonal subgroups, there is just \textit{one} out of
all 252 unordered partitions which works. The quiver is: 
\[
-C-C-W-W-H-W-H-W-H-W-H-H-
\]
with the ends identified. The chiral fermions in $(3,\bar{3},1)$ are shown
in Figure 1 which shows how three chiral families survive, arising from four
families and one anti-family.

This then is an explicit model with (i) $\mathcal{N}$=1, $SU(3)^{12}$
symmetry; (ii) successful predictivity for $\alpha _{3}(M_{Z})$ and $%
sin^{2}\theta (M_{Z})$ (see \cite{Frampton:2003cp}), (iii) three families; (iv) an
ameliorated gauge hierarchy where the ratio of the GUT scale to the weak
scale is less that 2 orders of magnitude rather that the greater than 12
orders of magnitude as in the MSSM \cite{ADFFL}.

To be complete, we should give the only alternative dodecagonal quiver with
similar properties. In it the C nodes are next-to-next-to-nearest neighbors
and the quiver is 
\[
-C-H-W-C-W-W-H-W-H-W-H-H-
\]
with the ends identified. These two models are in fact distinguishable at $%
M_{U}$ since the patterns of spontaneous symmetry breaking can be used to
reveal one of the two inequivalent embeddings. The symmetry breaking of the 
present model \cite{FK} can be summarized as follows.
We first let $(1,1,...,1,3,1,1...,1,\bar{3},1,1...,1)$ be defined as $(3,\bar{3}%
)_{i,j}$ for the $3$ and $\bar{3}$ in the $i^{th}$ and $j^{th}$ positions.
Then we can use $(3,\bar{3})_{1,2}$ to break$SU(3)\times SU(3)$ to $%
SU(3)_{C}.$ Similarly, we need $(3,\bar{3})_{3,4},$ $(3,\bar{3})_{4,6},$ $(3,%
\bar{3})_{6,8},$ and $(3,\bar{3})_{8,10}$ to break the W sector to $%
SU(3)_{L},$ and $(3,\bar{3})_{5,7},$ $(3,\bar{3})_{7,9},$ $(3,\bar{3}%
)_{9,11},$ and $(3,\bar{3})_{11,12}$ to break the H sector to $SU(3)_{R}.$
Next, VEVs for octets resulting from $(3,\bar{3})_{3,6}\rightarrow 1+8$ and $%
(3,\bar{3})_{9,12}\rightarrow 1+8$ can break the symmetry to $SU(3)_{C}%
\times SU(2)_{L}\times SU(2)_{R}\times U(1)_{L}\times U(1)_{R}.$ Finally, it
is easy to show that the appropriate doublets and charged singlets exist to
complete the breaking to $SU(3)_{C}\times SU(2)_{L}\times U(1)_{Y}.$ 


We conclude this section with some comments about the fermion mass spectrum.
The three families are chiral and so guaranteed to be light, 
while all other representations are vectorlike after
breaking to $SU(3) \times SU(2) \times U(1)$; 
and since only the three families are
protected by chiral symmetry, all other fermions should have masses
substantially higher than the weak scale. While this latter point is
likely to be true, we do not have a definitive proof that all the
vectorlike fermions are heavy, but it would require an accidental
symmetry to keep some of them light.  
Furthermore, even though the families are all light compared to
the unification scale,  the masses of the
three families arise from different VEVs in the pattern of spontaneous
symmetry breaking, as is easily seen from the quiver diagram. 
Thus in general, the masses should all be different, but
relatively light. The individual members of the families must also get
contributions to their masses in different ways, since again they come
from a variety of VEVs. Hence the model has sufficient robustness to
have a realistic mass spectrum.
Finally, although the Yukawa couplings are
initially fixed by conformality, soft conformal breaking terms which give
masses to fermions directly are not constrained. 
Since we are using an orbifolded N=3 theory, it is possible
that conformal invariance will not hold beyond first order. See, for
instance, \cite{Frampton:1999ti} where a large number of
orbifolded AdS/CFT models were investigated. These all have vanishing
one-loop beta functions, but somewhat less than ten percent of the
models are two-loop finite, hence we expect two-loop contributions to
lift any constraints on the Yukawa couplings. This along with the
discussion above makes for an interesting albeit complex fermion mass and
interaction structure. Mass terms for scalars are likewise unconstrained.


\bigskip \bigskip

\newpage

\noindent \textit{Cosmology and Summary}

\bigskip \bigskip

\noindent We have shown that there exists two $\mathcal{N}$=1 models that
unify at $\sim 4$ TeV. In the quiver diagram the two color nodes must be
nearest neighbors or next-to-next-to-nearest neighbors. For both choices the
assignment of all other nodes is dictated by the required symmetry breaking
pattern.

The motivation for supersymmetry is weakened by the amelioration of the GUT
gauge hierarchy but alternatively it may be justified by the concomitant
non-renormalization theorems which can translate one-loop conformality into
all-order conformality.

All $Z_{n}$ orbifolds have at least a $Z_{n}$ symmetry amongst the particle
irreps before the $SU(N)^{n}$ symmetry breaking. Breaking the gauge symmetry
also breaks the $Z_{n}$ and can lead to cosmic domain walls. Even though the
breaking scale is low ($\sim $ TeV), a single infinite wall is still
sufficiently massive to overclose the Universe if not inflated away. This
requires an inflation scale $<\phi >$ in the range $M_{W}\lesssim $ $<\phi >$
$\lesssim 4$ TeV with subsequent weak scale baryogenisis. A cosmologically
interesting possibility is for $Z_{n}$ to break before inflation, but for
the gauge group breaking to the standard model gauge group not to complete
until after inflation. Then cosmic strings and light monopoles can arise and
persist until the present. The monopoles\cite{Kephart:1995bi,Wick:2000yc}
would be light enough ($\sim $
100TeV) to avoid the cosmological monopole problem,
would also be ultra
relativistic with only electroweak interactions and be detectable at RICE.
One of the major historic reasons for resorting to string theory is that 
it can potentially provide a
consistant quantum gravity. While gravity is absent in the CFTs we have 
considered here, they
should be thought of as a sector of the full theory where gravity must 
enter. We are just beginning
to understand how gravity corrections come about,
so here all we
can do is just assume gravity arises in a natural way such that the 
effective theory below $\sim 4$ TeV
is ${\cal N}=1$ SUSY gauge theory plus general relativity \cite{Polchinski:2001tt}.

Baryon number generation is a concern for any model that requires inflation
at a low scale. Weak scale baryogenesis has now been ruled out for the
minimal standard model, but is still viable for the MSSM and its extensions
\cite{Dine:2003ax}. The fact that our models must inflate below 4 TeV means they
are testable cosmologically as well as testable and distinguishable at the
LHC. In analogy with the results of \cite{Frampton:2003cp}, it is easy to see
the properties for the two, similar but experimentially distinguishable,
models presented here are quite robust in the sense that fine tuning is not
necessary for them to agree with experimental particle physics data.

\bigskip
\bigskip
\bigskip
\bigskip

\noindent {\it Acknowledgements}

\bigskip

\noindent TWK thanks the Department of Physics and Astronomy at UNC Chapel
Hill and the Aspen Center for Physics,
and PHF thanks the Department of Physics and Astronomy at Vanderbilt
University for hospitality while this work was in progress. This work was
supported in part by the US Department of Energy under Grants No.
DE-FG02-97ER-41036 and No. DE-FG05-85ER40226.

\newpage

\bigskip
\bigskip

\noindent \underline{\bf Figure Caption}

\bigskip

\noindent Fig. 1.

\noindent  Chiral fermions transforming as $(3, \bar{3}, 1)$
in dodecagonal quiver diagram.
\bigskip

\newpage

\begin{figure}

\begin{center}

\epsfxsize=7.0in
\ \epsfbox{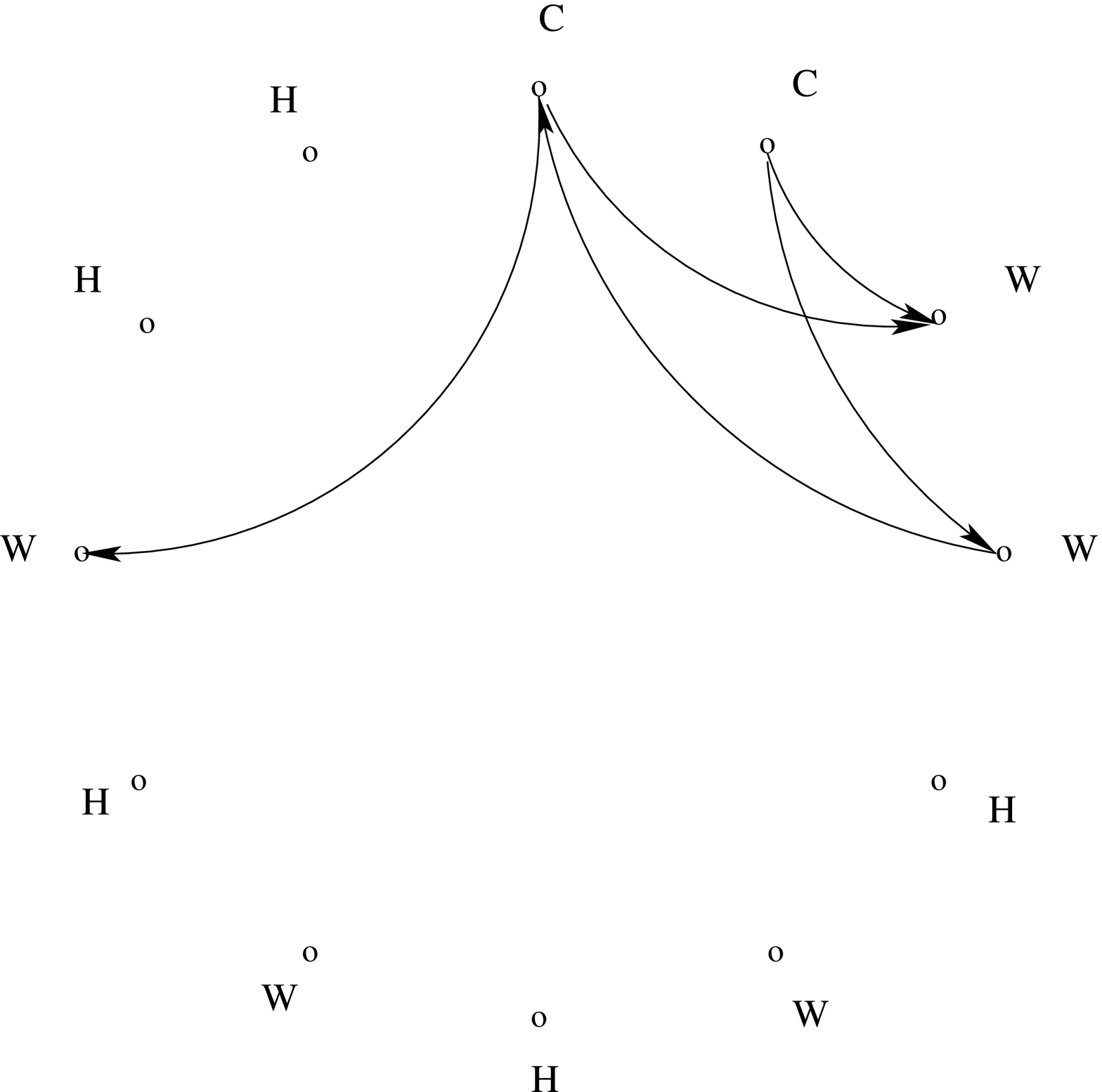}

\end{center}

\caption{}

\end{figure}

\end{document}